\documentclass[12pt]{article}
\usepackage[dvips]{graphicx}

\renewcommand{\bar}[1]{\overline{#1}}

\def\ru1{\rule[-0.4truecm]{0mm}{1truecm}}

\begin{document}
\newpage

\begin{flushright}
hep-ph/07XXXX \\
May  2007
\end{flushright}

\bigskip

{\centerline{\Large
\bf Single-Spin Asymmetries in Inclusive and}}
{\centerline{\Large \bf Exclusive Hadronic Processes}}

\vspace{22pt}

\centerline{
\bf Dae Sung Hwang$^{*}$, Jong Hyun Kim, and Seyong Kim$^{\dagger}$}

\vspace{8pt}
\centerline{Department of Physics, Sejong University, Seoul
143--747, Korea}

\centerline{e-mail: dshwang@sejong.ac.kr$^{*}$,
skim@sejong.ac.kr$^{\dagger}$}

\vspace*{1.0cm}



\begin{abstract}
\noindent
We investigate the single-spin asymmetries in inclusive and exclusive
processes of electron-proton scattering and electron-positron
annihilation. In the decomposition of hadronic tensor, a Lorentz
symmetric spin-dependent term is generally present. The existence of
such a term implies single-spin asymmetries in these processes and
these single-spin asymmetries for baryons can be understood in a
unified manner. We argue that it is important to measure the
single-spin asymmetries in both inclusive and exclusive processes for
the $\Lambda$ production at the present B-factories. This will lead to
the first measurement of the structure function of
the symmetric spin-dependent hadronic tensor
in these processes.
\end{abstract}

\vfill

\centerline{
}
\vfill


\newpage

\section{Introduction}

There have been important progresses in understanding the structure of
the proton recently. The Jeffereson Lab Hall A Collaboration
\cite{jones00} obtained the ratio of the proton's elastic
electromagnetic form factors $G_E/G_M$ by the measuring the
polarization transfers, whose result showed that the ratio decreases
systematically as $Q^2$ increases from 0.5 to 3.5 ${\rm GeV}^2$.  The
BABAR Collaboration \cite{BABAR06} measured, using the initial state
radiation, the $p{\bar p}$ mass dependence of the ratio of the
time-like proton electric and magnetic form factors and found that the ratio is significantly larger than 1 for masses up to 2.2
GeV. These new experimental results are more accurate than previous
ones and show the $q^2$ dependences of the form factors which are
different from old experimental results. These progresses are valuable
for improving our understanding of the nucleon structure and motivate
theoretical and experimental exploration on the hadronic structures. 

We are interested in the time-like nucleon form factors. They have
imaginary parts, differently from the space-like form factors, and
their complex phases can be measured by finding out the single-spin
asymmetry, ${\cal P}_y$, where ${\cal P}_y$ is defined as the
polarization of the produced nucleon in the direction normal to the
production plane. The dependence of ${\cal P}_y$ on the models of the
nucleon form factors were presented in \cite{BCHH04}. This single-spin
asymmetry may be measured at the future GSI $p{\bar p}$ experiment and
the up-graded DAFNE experiment. Also, one can measure such ${\cal
P}_y$ of produced $\Lambda$ particles at the present B-factories by
measuring the angular distribution of the daughter particles $p$ and
$\pi$ \cite{CGK07}. In this paper we interpret the single-spin
asymmetry, ${\cal P}_y$, in terms of the structure function of the
symmetric spin-dependent hadronic tensor, which has its origin in the
final-state strong-interaction phases. We call this structure function
$G_3$. Existence of $G_3$ permits single-spin asymmetries not only in
the exclusive processes but also in the inclusive processes without
violating the $T$-invariance in the time-like reactions since $G_3$
can be made from the final-state interactions in the time-like
reactions. If the single-spin asymmetry effect in the inclusive
$\Lambda$ production process in addition to the exclusive $\Lambda$
production processs is sizeable, its effect in both processes can be
measured at the present B-factories, and then the structure function
$G_3$ can be extracted. In particular, the results for $\Lambda$ will
be useful for the studies of the nucleon structure.  We can expect
that the sizes of these single-spin asymmetries are around 15 to 20 \%
since that of the process $pp\to \Lambda^{\uparrow}X$ is of that size
\cite{gourlay86}.

Symmetric spin-dependent structure function has been discussed in
various contexts previously. Christ and Lee \cite{CL66} proposed to
test the $T$-invariance by measuring the correlation function ${\bf
S}_{\rm in}\cdot ({\bf k}\times {\bf k}')$ in the inelastic scattering
$l^{\pm}+N\to l^{\pm}+\Gamma$, where ${\bf S}_{\rm in}$ is the
polarization vector of the initial nucleon $N$, ${\bf k}$ and ${\bf
k}'$ the initial and final momenta of the lepton $l^{\pm}$, $\Gamma$
all possible final states. They noted that this $T$-non-invariant
correlation by the single-photon exchange is independent of the sign
of the lepton charge, whereas such a $T$-invariant correlation by the
interference between the single-photon and the two-photon exchange
processes is proportional to the sign of the lepton charge. Christ and
Lee also showed that the unpolarized nucleon target can produce
spin-asymmetry. Gourdin \cite{gourdin72} introduced the symmetric
spin-dependent hadronic tensor in his study of the space-like deep
inelastic scattering of the longitudinally polarized leptons from
polarized nucleons. He noted that this symmetric spin-dependent
hadronic tensor measures a violation of the time reversal invariance.
These argument can be applied to time-like processes by
crossing and annihilation of lepton--anti-lepton pair into 
baryons can produce spin-polarized baryon. 

Unified understanding of single-spin asymmetries in electron-proton
scattering process and electron-positron annihilation process in terms
of symmetric spin-dependent structure function extends naturally
to inclusive processes as well because the structure function, $G_3$
is allowed by the Lorentz symmetry.
In this paper
we investigate the single-spin asymmetries in the inclusive and exclusive processes
of electron-proton scattering and electron-positron annihilation.
These single-spin asymmetries are presented in a unified manner through a symmetric
spin-dependent hadronic tensor.
It should be important to measure the single-spin asymmetries in both inclusive and
exclusive processes for the $\Lambda$ production at the present B-factories,
which will lead to the first measurement of the structure function of
the symmetric spin-dependent hadronic tensor.

\section{Space-like Processes}

\subsection{$e^-p\to e^-X$}

The structure of the proton is probed by the inelastic scattering
$e^-p\to e^-X$ in which the scattered electron is detected at a fixed
energy and angle, and $X$ indicates all possible states.
The structure functions are defined by
\begin{eqnarray}
W_{\mu\nu}&=&
\Bigl( -g_{\mu\nu}+{q_{\mu}q_{\nu}\over q^2}\Bigr)\ W_1(q^2,\nu )\ +\
{1\over M^2}\Big( P_\mu - {P\cdot q\over q^2}q_\mu \Big)
\Big( P_\nu - {P\cdot q\over q^2}q_\nu \Big)
\ W_2(q^2,\nu )
\nonumber\\
&+&{1\over M^3}\Big[
\Bigl( P_\mu -{P\cdot q\over q^2}q_\mu\Bigr)\epsilon_{\nu abc}s^aP^bq^c
+\Bigl( P_\nu -{P\cdot q\over q^2}q_\nu\Bigr)\epsilon_{\mu abc}s^aP^bq^c
\Big]G_3(q^2,\nu )\ ,
\label{w123}
\end{eqnarray}
in which $M$ is the proton mass and
$q=k-k'$ where $k$ and $k'$ are the momenta of the initial and
final electrons, respectively.
$M\nu =P\cdot q$ and we use $\epsilon_{0123}=1$.
$-q^2>0$ and $1<\omega\equiv {2M\nu \over |q^2|}<\infty$ for the space-like processes.

In (\ref{w123}) $W_1$ and $W_2$ are usual spin-independent structure functions,
and $G_3$ is a symmetric spin-dependent structure function which
measures a violation of the time reversal invariance in electromagnetic
interactions \cite{gourdin72}.
In this section we study the single-spin asymmetry which would be induced by $G_3$
in the space-like deep inelastic scattering.

Using the hadronic tensor $W_{\mu\nu}$ in (\ref{w123}) and
the symmetric part of the leptonic tensor given by
\begin{equation}
L^{\mu\nu (S)}=
2 \Bigl( k^\mu k'^\nu +k^\nu k'^\mu - k\cdot k' g^{\mu\nu}
\Bigr)\ ,
\label{h2space}
\end{equation}
we get the differential cross section in the rest frame of the target proton
\begin{eqnarray}
{d\sigma\over d\Omega dE'}
&=&{4\alpha^2E'^2\over |q^2|^2}\
{\rm sin}^2{\theta\over 2}
\Big[\ 2 W_1(q^2,\nu )\ +\ {\rm cot}^2{\theta\over 2}W_2(q^2,\nu )
\nonumber\\
&+&
({\vec s}\cdot {\hat y})\ {4(E+E') EE'\over M \, |q^2|}\ |{\rm sin}\theta |\
G_3(q^2,\nu )\ \Big]
\ ,
\label{w15}
\end{eqnarray}
where $E$ (${\hat k}$) and $E'$ (${\hat k'}$) are the initial and final energies
(momentum directions) and $\theta$ the scattering angle of the electron,
and ${\hat y}=({\hat k}\times {\hat k'})/|{\hat k}\times {\hat k'}|$.
From (\ref{w15}) the single-spin asymmetry is given by
\begin{equation}
{\cal P}_y={d\sigma ({\vec s}={\hat y})-d\sigma ({\vec s}=-{\hat y})\over
d\sigma ({\vec s}={\hat y})+d\sigma ({\vec s}=-{\hat y})}=
{{4(E+E') EE'\over M |q^2|}\ |{\rm sin}\theta |\ G_3(q^2,\nu )\over
2 W_1(q^2,\nu )\ +\ {\rm cot}^2{\theta\over 2}W_2(q^2,\nu )}\ .
\label{h2spaceinepy}
\end{equation}
Checking experimentally whether the above ${\cal P}_y$ is really zero
corresponds to the study by Christ and Lee on the $T$-violation
\cite{CL66}.


\subsection{$e^-p\rightarrow e^-p$}

The Dirac and
Pauli form factors $F_1(q^2)$ and $F_2(q^2)$ for a spin-${1\over 2}$
composite system are defined by
\begin{equation}
\langle P'| J^\mu (0) |P\rangle
= \bar u(P')\, \Big[\, F_1(q^2)\gamma^\mu +
F_2(q^2){i\over 2M}\sigma^{\mu\alpha}q_\alpha\, \Big] \, u(P)\ ,
\label{Drell1}
\end{equation}
where $u(P)$ is the bound state proton spinor and $q = P' -P$.
The electric and magnetic
form factors are defined by
\begin{equation}
G_E(q^2)=F_1(q^2)+{q^2\over 4M^2} F_2(q^2)\ ,\qquad 
G_M(q^2)=F_1(q^2)+ F_2(q^2)\ ,
\label{a3}
\end{equation}
where $G_E(0)=1$ and the proton magnetic moment is defined by
$\mu = {e\over 2M}G_M(0)$.

{}From Eqs. (\ref{Drell1}) and (\ref{a3}) we have the relations
\begin{eqnarray}
W_1^{\rm ex}(q^2,\nu )&=&-\ {q^2\over 4m^2}|G_M(q^2)|^2\ \delta (2M\nu +q^2)\ ,
\nonumber\\
W_2^{\rm ex}(q^2,\nu )&=&{|G_E(q^2)|^2-{q^2\over 4M^2}|G_M(q^2)|^2\over 1-{q^2\over 4M^2}}
\ \delta (2M\nu +q^2)\ ,
\nonumber\\
G_3^{\rm ex}(q^2,\nu )&=&-\ {{\rm Im}[G_M^*G_E]\over 2\Big( 1-{q^2\over 4M^2}\Big) }
\ \delta (2M\nu +q^2)\ ,
\label{space11}
\end{eqnarray}
and then, we get the differential cross section given by
\begin{eqnarray}
{d\sigma\over d{\rm cos}\theta}
&=&
{\pi {\alpha}^2E'{\rm cos}^2{\theta\over 2}\over 2E^3{\rm sin}^4{\theta\over 2}}
\Big( {1\over 1+\tau}\Big)\ 
\Big[\ |G_E(q^2)|^2+{\tau\over \epsilon}|G_M(q^2)|^2
\nonumber\\
&-&
({\vec s}\cdot {\hat y})\ {E+E'\over M}|{\rm tan}{\theta\over 2}|\
{\rm Im}[G_M^*G_E]\ \Big]
\ ,
\label{w15EX}
\end{eqnarray}
where ${\hat y}=({\hat k}\times {\hat k'})/|{\hat k}\times {\hat k'}|$,
$\tau ={|q^2|\over 4M^2}$, and
${1 / \epsilon}=1+2(1+\tau ){\rm tan}^2{\theta\over 2}$.
The scattering angle dependences of $\tau$ and $E'$ are given by
$\tau=
\Big( ({E\over M})^2{\rm sin}^2{\theta\over 2}\Big) / \Big( 1+2{E\over M}
{\rm sin}^2{\theta\over 2}\Big)$ and
$E'=E / \Big( 1+2{E\over M}{\rm sin}^2{\theta\over 2}\Big)$.
From (\ref{w15EX}) the single-spin asymmetry is given by
\begin{equation}
{\cal P}_y=
{-\ {E+E'\over M}\ |{\rm tan}{\theta\over 2}|\ {\rm Im}[G_M^*G_E]\over
|G_E(q^2)|^2+{\tau\over \epsilon}|G_M(q^2)|^2}\ .
\label{h2spacEXpy}
\end{equation}
In reality, the above ${\cal P}_y$ is expected to be zero since the
Hermiticity of the electromagnetic current renders the space-like form
factors real. However, it is useful to check experimentally whether
${\cal P}_y$ in Eq.(\ref{h2spacEXpy}) really vanishes.


\section{Time-like Processes}

\subsection{$e^+e^-\to pX$}


The structure of the proton is probed by the inelastic scattering
$e^+e^-\to pX$ in which the produced proton is detected at a fixed
energy and angle, and $X$ indicates all possible states.
The structure functions are defined by 
\begin{eqnarray}
{\bar{W}}_{\mu\nu}&=&
\Bigl( -g_{\mu\nu}+{q_{\mu}q_{\nu}\over q^2}\Bigr)\ {\bar{W}}_1(q^2,\nu )
\ +\
{1\over M^2}\Big( P_\mu - {P\cdot q\over q^2}q_\mu \Big)
\Big( P_\nu - {P\cdot q\over q^2}q_\nu \Big)
{\bar{W}}_2(q^2,\nu ) \ \ \ \ \
\nonumber\\
&+&{1\over M^3}\Big[
\Bigl( P_\mu -{P\cdot q\over q^2}q_\mu\Bigr)\epsilon_{\nu abc}s^aP^bq^c
+\Bigl( P_\nu -{P\cdot q\over q^2}q_\nu\Bigr)\epsilon_{\mu abc}s^aP^bq^c
\Big]\ G_3(q^2,\nu )\ ,
\label{w123time}
\end{eqnarray}
in which $M$ is the proton mass and
$q=k+k'$ where $k$ and $k'$ are the momenta of the initial electron and
positron, respectively.
$M\nu =P\cdot q$ and we use $\epsilon_{0123}=1$.
$q^2>0$ and $0<\omega\equiv {2M\nu \over q^2}<1$ for the time-like processes.

In (\ref{w123time}) ${\bar{W}}_1$ and ${\bar{W}}_2$ are useal spin-independent
structure functions \cite{DLY1},
and $G_3$ is a symmetric spin-dependent structure function which
has its origin in the strong-interaction phases \cite{gourdin72, gatto74}.
In this section we study the single-spin asymmetry which would be induced by $G_3$
in the time-like deep inelatic scattering.

Using the hadronic tensor $W_{\mu\nu}$ in (\ref{w123}) and
the symmetric part of the leptonic tensor given by
\begin{equation}
L^{\mu\nu (S)}=
2 \Bigl( k^\mu k'^\nu +k^\nu k'^\mu - k\cdot k' g^{\mu\nu}
\Bigr)\ ,
\label{h2}
\end{equation}
we get the differential cross section in the center of mass frame of the
initial electron and positron
\begin{eqnarray}
&&{d^2\sigma\over dE d{\cos{\theta}}}
\nonumber\\
&=&{2\pi\alpha^2M\over (q^2)^{3\over 2}}
\Big( {2M\nu \over q^2}\Big)
\Big(1-{q^2\over \nu^2}{\Big)}^{1\over 2} \
\Big(\
\Big[ 2{\bar{W}}_1(q^2,\nu ) \ +\
\Big( {2M\nu \over q^2}\Big) \Big( 1 - {q^2\over \nu^2} \Big)\
{\rm sin}^2{\theta}\
{\nu {\bar{W}}_2(q^2,\nu ) \over 2M}
\Big]
\nonumber\\
&+&
({\vec s}\cdot {\hat y})
\ {(q^2)^{3\over 2}\over 4M^3}\ \Big( {2M\nu \over q^2} {\Big)}^2\
\Big( 1 - {q^2\over \nu^2} \Big)\ 
{\rm sin}2\theta \ G_3(q^2,\nu )
\ \Big) 
\ ,
\label{w15timea}
\end{eqnarray}
where $E$ is the energy of the produced proton, 
${\hat k}$ and ${\hat p}$ the initial electron and produced proton
momentum directions, and $\theta$ the angle between ${\hat k}$ and ${\hat p}$,
and ${\hat y}=({\hat k}\times {\hat p})/|{\hat k}\times {\hat p}|$.
From (\ref{w15timea}) the single-spin asymmetry is given by
\begin{equation}
{\cal P}_y=
{{(q^2)^{3\over 2}\over 4M^3}\ \Big( {2M\nu \over q^2} {\Big)}^2\
\Big( 1 - {q^2\over \nu^2} \Big)\
{\rm sin}2\theta \ G_3(q^2,\nu )\over
2{\bar{W}}_1(q^2,\nu ) \ +\
\Big( {2M\nu \over q^2}\Big) \Big( 1 - {q^2\over \nu^2} \Big)\
{\rm sin}^2{\theta}\
{\nu {\bar{W}}_2(q^2,\nu ) \over 2M}}
\ .
\label{h2timeinepy}
\end{equation}

We also have
\begin{equation}
{1\over 2}\Big( {d^2\sigma ({\vec s}={\hat y})\over dE d{\cos{\theta}}}
-{d^2\sigma ({\vec s}=-{\hat y})\over dE d{\cos{\theta}}}\Big)\ =\
{\pi\alpha^2\over 2M^2}\ 
\Big( {2M\nu \over q^2} {\Big)}^3\
\Big( 1 - {q^2\over \nu^2} {\Big)}^{3\over 2}\ 
{\rm sin}2\theta \ G_3(q^2,\nu )
\ .
\label{w15timeb}
\end{equation}


In the Bjorken limit of the deep inelastic process
\cite{DLY1,bjorken69}, we have
\begin{equation}
M{\bar W}_1=-F_1(\omega )\ ,\qquad
\nu{\bar W}_2=F_2(\omega )\ ,\qquad {\rm and}\ \
F_1(\omega )={\omega\over 2}F_2(\omega )\ .
\label{cgrelationtime}
\end{equation}
Then, the differential cross section is given by
\begin{eqnarray}
{d^2\sigma\over dE d{\cos{\theta}}}
&=&{2\pi\alpha^2M\over (q^2)^{3\over 2}}\
\omega\
\Big(1-{4M^2\over q^2}{1\over \omega^2}{\Big)}^{1\over 2} \
\nonumber\\
&\times&
\Big(\
{2\over M}\ \Big( -F_1(\omega )\Big)\
\Big[ 1 \ -\
{1\over 2}\ \Big( 1 - {4M^2\over q^2}{1\over \omega^2} \Big)
\ {\rm sin}^2{\theta}\Big]
\nonumber\\
&&+\ ({\vec s}\cdot {\hat y})\
{(q^2)^{3\over 2}\over 4M^3}\ \omega^2\
\Big( 1 - {4M^2\over q^2}{1\over \omega^2} \Big)\
{\rm sin}2\theta \ G_3(\omega )
\ \Big)
\ ,
\label{w15timeb}
\end{eqnarray}
and the single-spin asymmetry is given by
\begin{equation}
{\cal P}_y\ =\
{G_3(\omega )\ {(q^2)^{3\over 2}\over 4M^3}\ \omega^2\
\Big( 1 - {4M^2\over q^2}{1\over \omega^2} \Big)\
\over
\Big( -F_1(\omega )\Big)\ {2\over M}}\
{{\rm sin}2\theta 
\over
\Big[ 1 \ -\
{1\over 2}\ \Big( 1 - {4M^2\over q^2}{1\over \omega^2} \Big)
\ {\rm sin}^2{\theta}\Big]}
\ .
\label{pytimeinc}
\end{equation}
We present the graph of Eq. (\ref{pytimeinc}) in Fig. 1.

\begin{figure*}[ht]
\vskip 2.0cm
\begin{center}
\includegraphics[width=5.6in]{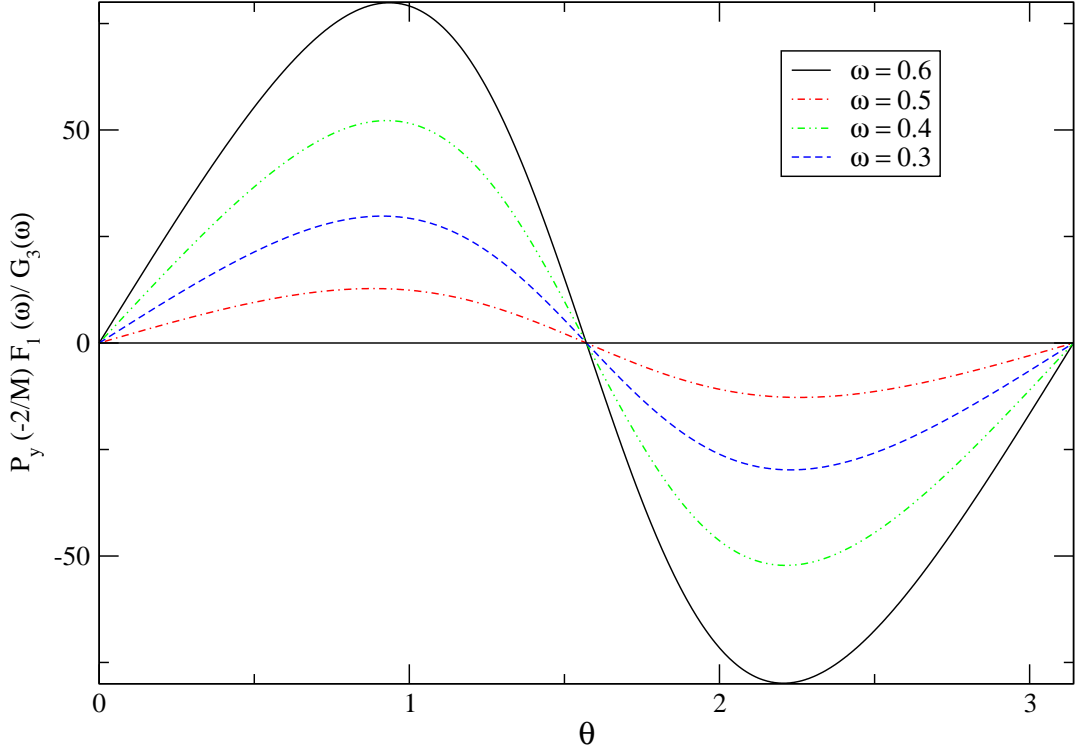}
\vskip 0.5cm
\caption{single-spin asymmetry, ${\cal P}_y \; (-\frac{2}{M})\; \frac{F_1
    (\omega)}{G_3(\omega)}$ from Eq. (\ref{pytimeinc}) }
\label{hadronmass}
\end{center}
\setcounter{figure}{1}
\end{figure*}


\subsection{$e^-e^+\rightarrow p{\bar{p}}$}

The Dirac and
Pauli form factors $F_1(q^2)$ and $F_2(q^2)$ for a spin-${1\over 2}$
composite system are defined by
\begin{equation}
\langle pp'| J^\mu (0) |0\rangle
= \bar u(P)\, \Big[\, F_1(q^2)\gamma^\mu +
F_2(q^2){i\over 2M}\sigma^{\mu\alpha}q_\alpha\, \Big] \, v(P')
\ ,
\label{c1timee}
\end{equation}
where $u(P)$ and $v(P')$ are bound state proton and anti-proton spinors,
where $q = P' +P$.
The electric and magnetic
form factors, $G_E(q^2)$ and $G_M(q^2)$, are defined by
\begin{equation}
G_E(q^2)=F_1(q^2) + {q^2\over 4m^2} F_2(q^2)\ ,\qquad
G_M(q^2)=F_1(q^2) + F_2(q^2)\ .
\label{c3timee}
\end{equation}
{}From Eqs. (\ref{Drell1}) and (\ref{a3}) we have the relations
\begin{eqnarray}
{\bar{W}}_1^{\rm ex}(q^2,\nu )&=&{q^2\over 4M^2}|G_M(q^2)|^2
\ \delta (2M\nu -q^2)\ ,
\nonumber\\
{\bar{W}}_2^{\rm ex}(q^2,\nu )&=&-\ {|G_E(q^2)|^2-{q^2\over 4M^2}|G_M(q^2)|^2\over 1-{q^2\over 4M^2}}
\ \delta (2M\nu -q^2)\ ,
\nonumber\\
G_3^{\rm ex}(q^2,\nu )&=&{{\rm Im}[G_M^*G_E]\over 2\Big( 1-{q^2\over 4M^2}\Big)}
\ \delta (2M\nu -q^2)\ .
\label{timee12}
\end{eqnarray}
and then, we get the differential cross section given by
\begin{eqnarray}
{d\sigma\over d{\rm cos}\theta}
&=&
{\pi {\alpha}^2{\sqrt{1-{4M^2\over q^2}}}\over 2q^2}\
\Big[\ |G_M(q^2)|^2(1+{\rm cos}^2\theta )+{4M^2\over q^2}|G_E(q^2)|^2{\rm sin}^2\theta
\nonumber\\
&&-\ ({\vec s}\cdot {\hat y})\ {2M\over {\sqrt{q^2}}}\ {\rm sin}2\theta \
{\rm Im}[G_M^*G_E]
\ \Big] 
\ ,
\label{crosecpp}
\end{eqnarray}
where ${\hat y}=({\hat k}\times {\hat p}) / |{\hat k}\times {\hat p}|$.
From (\ref{w15EX}) the single-spin asymmetry is given by \cite{BCHH04}
\begin{equation}
{\cal P}_y=
{-\ {2M\over {\sqrt{q^2}}}\ {\rm sin}2\theta \ {\rm Im}[G_M^*G_E]\over
|G_M(q^2)|^2(1+{\rm cos}^2\theta )+{4M^2\over q^2}|G_E(q^2)|^2{\rm sin}^2\theta }
\ .
\label{crosecpp}
\end{equation}


\section{Conclusion}

A symmetric spin-dependent structure function, $G_3 (q^2,
\nu)$, is allowed in the decomposition of hadronic tensor
and this structure function can exist in the time-like processes of
electron-positron annihilation into hadrons from the strong phases
which are made by the final-state interactions. 

By use of this symmetric spin-dependent structure function, we present
single-spin asymmetries in inclusive and exclusive processes involving
baryons in the final state in a unified way and strongly suggests that
single-spin asymmetries are generic phenomena. The existence of $G_3
(q^2, \nu)$ can be measured by $\sin 2 \theta $ distribution
in the differential cross-section given in Eq. (\ref{w15timea}).

We argue that it is important to measure the
single-spin asymmetries in both inclusive and exclusive processes for
the $\Lambda$ production at the present B-factories. This will lead to
the first measurement of the structure function of
the symmetric spin-dependent hadronic tensor
in these processes.

\section*{Acknowledgments}
This work was supported in part by the International Cooperation
Program of the KICOS (Korea Foundation for International Cooperation
of Science \& Technology).

\end{document}